\documentclass[sigconf]{acmart}

\usepackage{comment}
\usepackage{enumitem}
\usepackage{blindtext}
\usepackage{circledtext}
\usepackage{pifont}
\usepackage{pgffor}
\usepackage{tikz}
\usepackage{graphicx}
\usepackage{lipsum}

\usepackage{listings}
\usepackage{xcolor}
\definecolor{codegreen}{rgb}{0,0.6,0}
\definecolor{codegray}{rgb}{0.5,0.5,0.5}
\definecolor{codepurple}{rgb}{0.58,0,0.82}
\definecolor{backcolour}{rgb}{0.95,0.95,0.92}

\lstdefinestyle{mystyle}{
    backgroundcolor=\color{backcolour},   
    commentstyle=\color{codegreen},
    keywordstyle=\color{magenta},
    numberstyle=\tiny\color{codegray},
    stringstyle=\color{codepurple},
    basicstyle=\ttfamily\footnotesize,
    breakatwhitespace=false,         
    breaklines=true,                 
    captionpos=b,                    
    keepspaces=true,                 
    numbers=none,
    numbersep=5pt,                  
    showspaces=false,                
    showstringspaces=false,
    showtabs=false,                  
    tabsize=2
}

\AtBeginDocument{%
  }

\setcopyright{acmcopyright}
\copyrightyear{2024}
\acmYear{2024}
\acmDOI{XXXXXXX.XXXXXXX}

\renewcommand\footnotetextcopyrightpermission[1]{} 
\begin{document}
\title{Makinote: An FPGA-Based HW/SW  Platform for \\Pre-Silicon Emulation of RISC-V Designs}

%



\author{Elias Perdomo}
\affiliation{%
  \institution{Barcelona Supercomputing Center}  
  \country{Spain}
  }
\affiliation{%
\institution{Universitat Polit\`ecnica de Catalunya}
  \country{Spain}
  }

\author{Alexander Kropotov}
\affiliation{%
  \institution{Barcelona Supercomputing Center}  
  \country{Spain}
  }

\author{Francelly Cano}
\affiliation{%
  \institution{Barcelona Supercomputing Center}  
  \country{Spain}
  }

\author{Syed Zafar}
\affiliation{%
  \institution{Barcelona Supercomputing Center}  
  \country{Spain}
  }

\author{Teresa Cervero}
\affiliation{%
  \institution{Barcelona Supercomputing Center}  
  \country{Spain}
  }

\author{Xavier Martorell}
\affiliation{%
  \institution{Barcelona Supercomputing Center}  
  \country{Spain}
  }
\affiliation{%
\institution{Universitat Polit\`ecnica de Catalunya}
  \country{Spain}
  }

\author{Behzad Salami}
\affiliation{%
  \institution{Barcelona Supercomputing Center}  
  \country{Spain}
  }

\renewcommand{\shortauthors}{}

\begin{abstract}
Emulating chip functionality before silicon production is crucial, especially with the increasing prevalence of RISC-V-based designs. FPGAs are promising candidates for such purposes due to their high-speed and reconfigurable architecture. In this paper, we introduce our \textbf{\textit{Makinote}}, an FPGA-based Cluster platform, hosted at Barcelona Supercomputing Center (BSC-CNS), which is composed of a large number of FPGAs (in total 96 AMD/Xilinx Alveo U55c) to emulate massive size RTL designs (up to 750M ASIC cells)\footnote{Our hardware platform is initially targeted for research purpose and under collaboration agreements it can get available for external researchers}. In addition, we introduce our FPGA shell as a powerful tool to facilitate the utilization of such a large FPGA cluster with minimal effort needed by the designers. The proposed FPGA shell provides an easy-to-use interface for the RTL developers to rapidly port such design into several FPGAs by automatically connecting to the necessary ports, e.g., PCIe Gen4, DRAM (DDR4 and HBM), ETH10g/100g. Moreover, specific drivers for exploiting RISC-V based architectures are provided within the set of tools associated with the FPGA shell. We release the tool online for further extensions.\footnote{\url{https://github.com/MEEPproject/fpga_shell} }

We validate the efficiency of our hardware platform (i.e., FPGA cluster) and the software tool (i.e., FPGA Shell) by emulating a RISC-V processor and experimenting HPC Challenge application running on 32 FPGAs. Our results demonstrate that the performance improves by 8 times over the single-FPGA case. 

\end{abstract}


\sloppy
\maketitle

\section{Introduction}
Recently, logic emulation emerged as the dominant approach addressing chip emulation challenges and its design explosion problems.
These scalability problems arise from the increasing complexity of current processors and Systems-on-Chip (SoCs), which integrate hundreds of complex components, such as cores, accelerators, and memory hierarchies.
Simulating such designs on conventional systems is notably time-consuming.
Traditional simulation can be accelerated using emulation systems, where FPGAs on emulation boards replace a portion or the entire design, with the remaining simulation tasks handled by software on the host computer.

\begin{figure}[!b]
\centerline{\includegraphics[width=\linewidth]{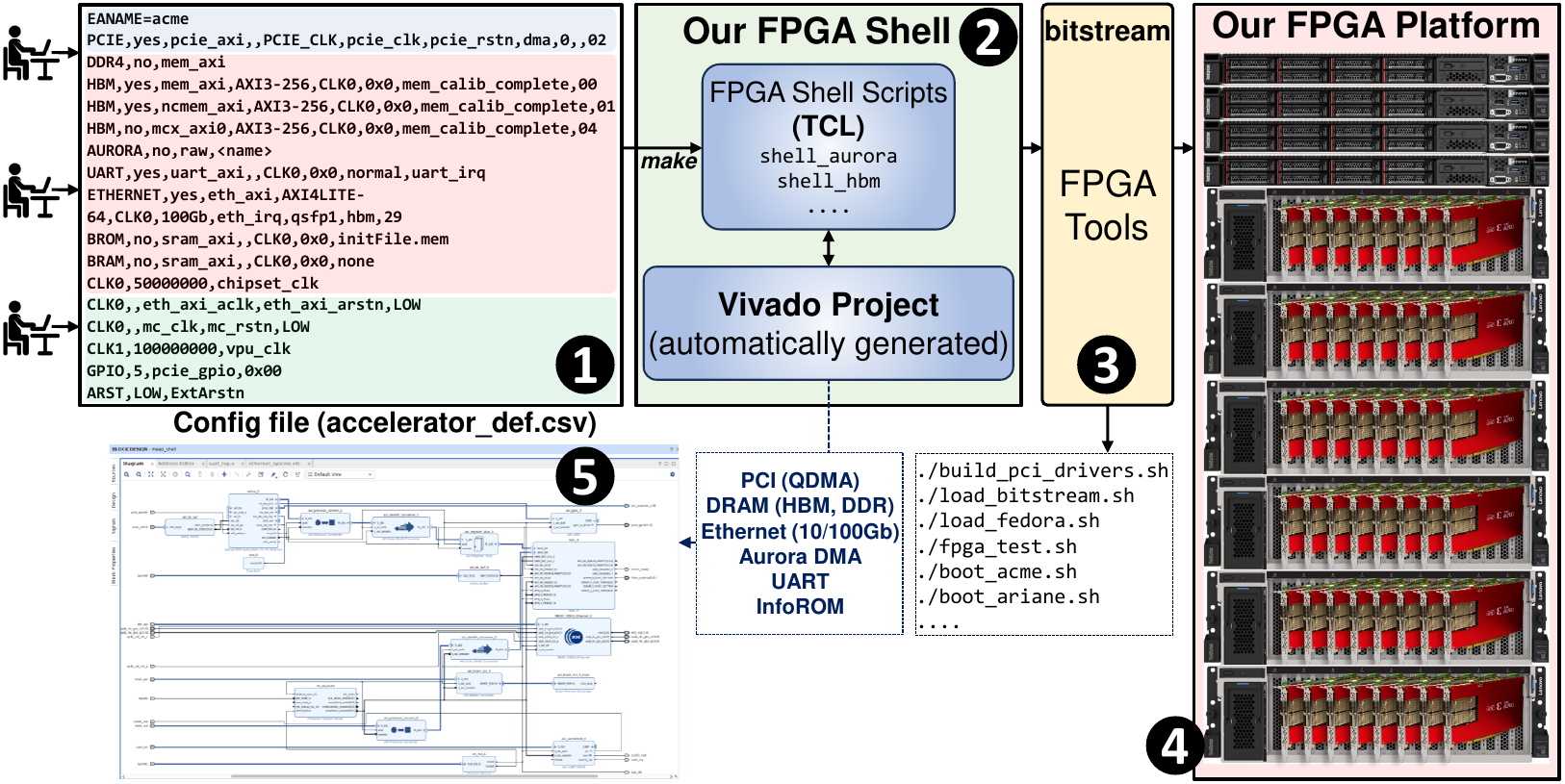}}
\vspace{-0.3cm}
\caption{Our hardware/software FPGA-based platform including FPGA Shell and FPGA Cluster}
\label{fig:fig_rapid_prototyping}
\end{figure}

\begin{figure*}[h]
    \centerline{\includegraphics[width=0.9\textwidth]{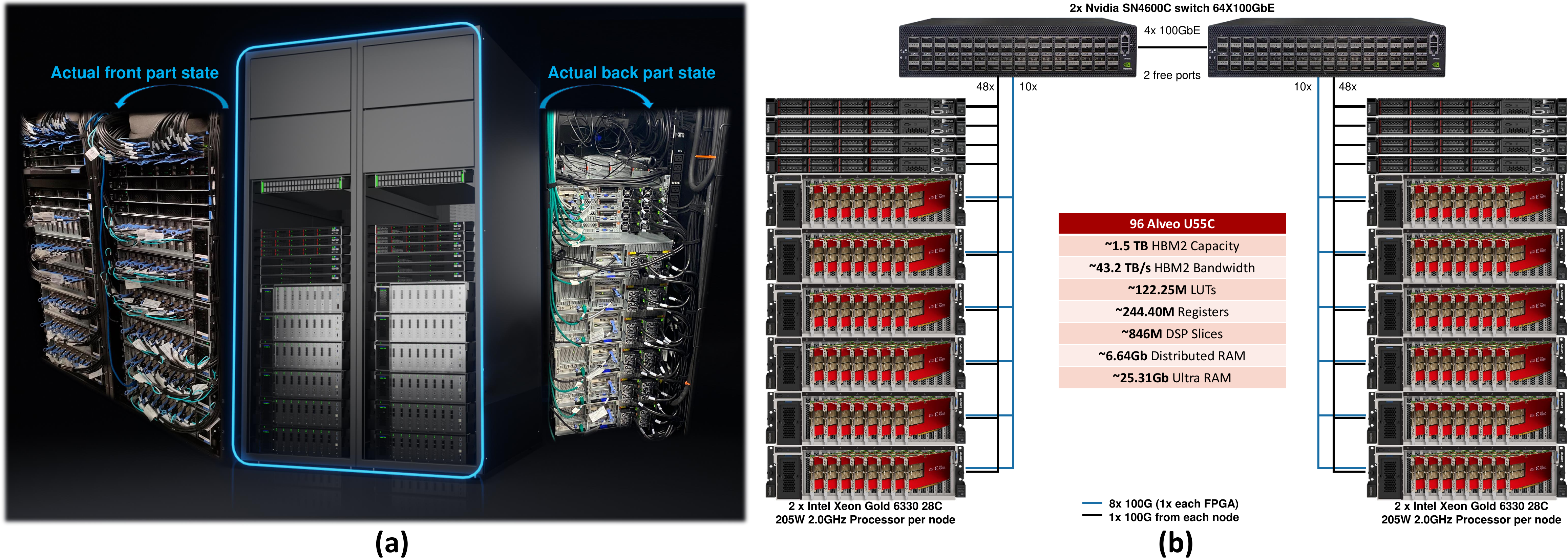}}
    \vspace{-0.4cm}
    \caption{FPGA Cluster (a) Physical view and (b) Schematic view.}
    \label{fig:cluster-detailed}
\end{figure*}

FPGA-based logic emulation foundational concept was described in \cite{babb_logic_1997}.
A logic emulation system consists of one or more emulation boards with single or multiple FPGAs and other integrated circuits (ICs) such as memories \cite{tessier_multi-fpga_2008,hosokawa_efficient_2007}.
The notion of using multiple FPGAs for improving logic emulation was proposed in \cite{hauck_software_1996} as a reconfigurable computing application \cite{compton_reconfigurable_2002, hauck_reconfigurable_2010}. 
To achieve high-speed emulation, a transaction-based approach is commonly employed, \cite{borgstrom_transaction-based_2014,hassoun_transaction-based_2005} allowing emulation and computer software to execute independently, minimizing waiting times.
The hardware for logic emulation involves three main types of chips\cite{hung_challenges_2018}: (1) customized processor,
(2) customized Field Programmable Gate Array (FPGA) and processor
and (3) commercial FPGAs.
Whereas (1) and (2) are application-specific integrated circuits (ASICs), (3) is not.
Each has its advantages and drawbacks, as discussed in detail in \cite{bailey_emulator_2010,rizzatti_whats_2014}.

The growing capabilities of FPGAs introduce compelling emulation-based alternatives.
Leveraging their inherently programmable architecture and hardware-speed executions enables combining fast simulation times with high accuracy and flexibility \cite{dave_implementing_2006, wawrzynek_ramp_2007}.
FPGA emulation can significantly reduce the simulation time of large-scale design from years to days  \cite{elsabbagh_accelerating_2023}.
For one trillion cycles (\textasciitilde20 minutes) i.e., this extensive cycle count decreased from an estimated 3 years (1095 days) to a mere 8.7 days.

Reduced Instruction Set Computer (RISC-V) based processors and SoCs are current complex systems that benefit from FPGA emulation capabilities. 
RISC-V is an open standard for an instruction set architecture that has garnered substantial momentum over the past decade \cite{risc-v_international_specification_nodate}.
Industry benefits from this open and customizable architecture, promoting innovation, reducing vendor lock-in, and enabling cost-effective development of specialized processors.
Additionally, it empowers academia to explore innovative computer architecture with a relevant, accessible standard, providing a pathway toward real-world commercial usage \cite{cui_risc-v_2023,mezger_survey_2022}.

Pre-silicon emulation using FPGAs has a long history, both academia (e.g., LiTEX\cite{litex_github_2023}, FireSim\cite{firesim_powered_by_jekyll__minimal_mistakes_firesim_2023}) and industry (e.g., Synopsys HAPS\cite{synopsys_haps-100_2023}, Cadence Palladium \cite{cadence_design_systems_palladium_2023}, Siemens Veloce\cite{siemens_veloce_2023}) have put significant effort into it.
Although commercial platforms propose a complete hardware and software solution, they are not easily available for research purposes.
On the other side, open-source platforms do not provide a complete hardware and software solution and have limited features.
In contrast, we propose a complete hardware and software solution initially targeted for research purposes; our hardware platform is built based on commercial-off-the-shelf (COTS) FPGAs and our software platform is built on top of standard AMD/Xilinx IPs.


In this paper, we present an FPGA-Based Hardware/Software platform for pre-silicon emulation of
RISC-V designs.  \textit{Our Hardware Platform} is a large-scale FPGA-based cluster comprising 96 AMD/Xilinx Alveo U55c boards and provides the capability of emulating large RISC-V designs, up to 730M ASIC cells. In addition, \textit{Our Software Platform}, i.e., the FPGA shell, is a configurable, scalable, extensible, and easy-to-use tool to facilitate using the cluster by enabling the widely-used ports of FPGA (e.g., PCIe, Ethernet, Aurora, HBM).

Fig.\ref{fig:fig_rapid_prototyping} shows the overall flow to work with our platforms. As a first step, through a configuration file, the users enable and configure the interfaces of the FPGA Shell need based on their design requirements \ding{182}. Our FPGA Shell \ding{183} depending on the file options,  enables the desired macros and IPs and automatically generates a Vivado project connecting the user design to the desired interfaces \ding{184} and generates a bitstream of the interfaced design. Once the bitstream is generated, by using our FPGA tools \ding{185} all the needed drivers and the bitstream can be loaded in our cluster \ding{186}.

\section{\textit{Makinote:} A Large-Scale FPGA-based Cluster}
\label{sec:hw_description}

\textit{Our HW Platform} is a large-scale multi-FPGA cluster designed to support hardware/software co-design activities and provides the capability of emulating large RISC-V designs, up to 730M ASIC cells.
Our cluster empowers applications to run at a system level with multiple nodes and FPGAs per node.
It also provides the foundational infrastructure for simulation, emulation, and software development to support the RISC-V ecosystem. In Fig.\ref{fig:cluster-detailed}\textbf{(a)} a Physical view of our cluster is illustrated.

\textit{Our HW Platform} comprises 96 Xilinx Alveo U55C Data Center Accelerator Card, from AMD \cite{amd_xilinx_alveo_2022} targeting Virtex XCU55 UltraScale+ HBM high-density FPGA modules. 
The designation of the FPGA used on the Alveo U55C board is the XCU55C, a multi-die FPGA made of XCVU33P + two XCVU11P die, which is rebadged XCVU37P \cite{amd_xilinx_ultrascale_2023}.
The Alveo U55C has also various hard IP blocks, including HBM and memory controllers, PCIe controllers and drivers as well as QSFP and Ethernet functionality.
These hard macros establish the primary functionalities required for numerous future designs to be emulated by an FPGA-based platform.

\begin{figure}[!b]
\centerline{\includegraphics[width=\linewidth]{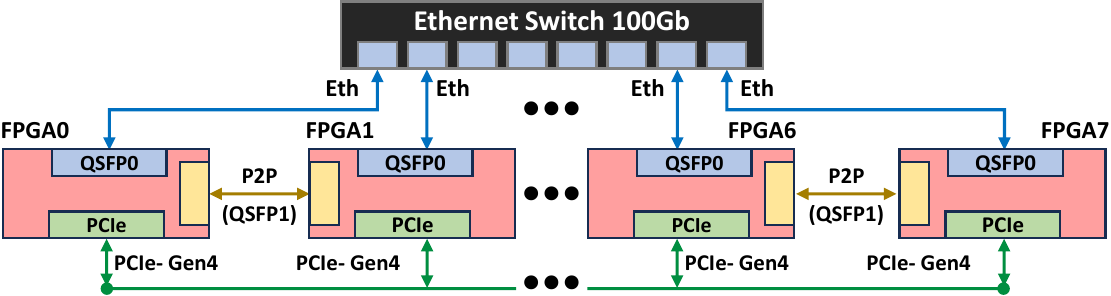}}
\vspace{-0.4cm}
\caption{The interconnection network of FPGAs in the cluster (illustrated for a single node)}
\label{fig:fig_ethernet_aurora_switch_pastel_vec}
\vspace{-0.4cm}
\end{figure}

Our cluster is composed of 12 nodes, hosting 96 Alveo U55C cards (8 cards per node), and divided into two identical racks (6 nodes per rack). 
In Fig.\ref{fig:cluster-detailed}\textbf{(b)} a detailed schematic view of the infrastructure is presented.
The interconnection among FPGAs is based on QSFP+ ports of the FPGAs (two ports per card) as well as PCIe Gen4. The cluster is equipped with two 64-port 100gbs Ethernet switches, which are used to connect cards through QSFP1. As a soft configuration, we use the other QSFP0 port of the FPGA cards to establish a direct FPGA-to-FPGA connection among cards in pairs.
Furthermore, within the nodes, all 8 FPGAs are connected through PCIe Gen4 for the host-FPGA communication.
The all-to-all communication provides flexibility for interconnecting FPGAs together.
Fig.~\ref{fig:fig_ethernet_aurora_switch_pastel_vec} showcases these interface details within a node.

\subsection{\textit{\textbf{Makinote}:} access methodology}

When accessing our FPGA-based cluster is crucial to know how it is organized as well as the methodology to be followed to be able to place our designs successfully.
Currently, our cluster has distinct types of nodes, each serving a specific purpose:
\begin{enumerate}
  \item \textbf{Management nodes (4 nodes):} One of these is the \textit{Login node} which is the entry point for accessing the FPGA cluster and operating one or more FPGA cards using SLURM for resource allocation. The remaining 3 nodes are exclusively for overall rack operation control.
  \item \textbf{Compute nodes (4 nodes):} General-purpose compute nodes that are not associated with FPGA usage.
  \item \textbf{FPGA node (12 nodes):} Each has eight FPGAs.
  Users can allocate 1 or many FPGAs in a node or multiple nodes.
\end{enumerate}

\begin{figure}[!t]
\centerline{\includegraphics[width=\linewidth]{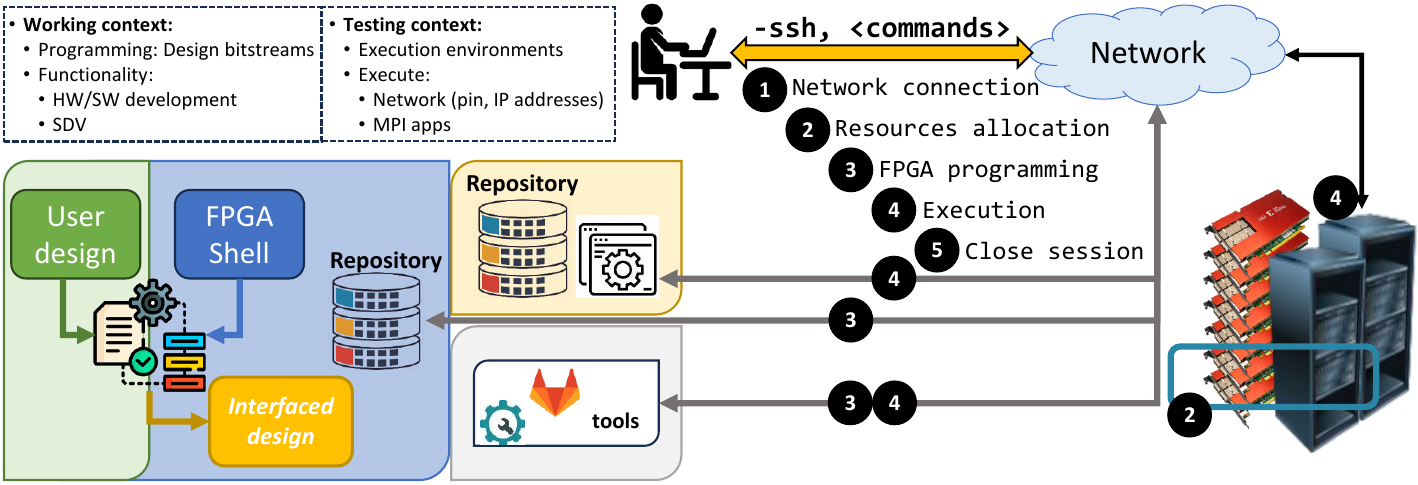}}
\vspace{-0.4cm}
\caption{General schematic of the network designs and interactions.}
\label{fig:fig_access_cluster_vec}
\end{figure}

An overview of the large-scale FPGA machine's organization at the network level and the interactions amongst different components is illustrated in Fig.~\ref{fig:fig_access_cluster_vec}.
This diagram represents the final system and the steps to access the numerous resources available:
\begin{itemize}[align=right,itemindent=1em,labelsep=2pt,labelwidth=1em,leftmargin=*,nosep]
   \item The first step is to stablish a connection to our large-scale FPGA cluster through the Login node via SSH \ding{182}.  Once logged, resources must be allocated and configured via SLURM.
   \item Then, if there is availability, the system : 1) allocates the requested resources, 2) configures the resources, and 3) deploys the system and provides access to the user \ding{183}.
   \item A bitstream of the user design is "wrapped up" by our FPGA Shell (further explained in the following section) using the available repositories to get access to the different available interfaces (e.g., PCIe, Ethernet, Aurora, HBM)\ding{184}.
   \item Finally, using a set of tools we developed the required drivers and the bitstream can be programmed on the cluster \ding{185}.
  \item Once a user finishes, the session is guaranteed to be properly closed and user data is removed from the system \ding{186}..
 \end{itemize}

The current system configuration, demands users to specify two parameters as part of the connection process: 
\begin{itemize}[align=right,itemindent=1em,labelsep=2pt,labelwidth=1em,leftmargin=*,nosep]
   \item All FPGA cards are configured with a specified Ethernet speed: 10Gb Ethernet ("10g"), or 100Gb Ethernet ("100g"). The latter is also a default configuration ("auto").
   \item PCIe configuration for its Direct Memory Access (DMA) kernel module: "none", "xdma" or "qdma".
 \end{itemize}
These constraints can be applied individually or in combination, depending on the specific user needs.

\section{\textit{Our SW Platform:} FPGA Shell Enabling Seamless FPGA Utilization}
\label{sec:sw_description}

\textit{Our SW platform}
simplifies FPGAs usability, enabling expert and non-expert users to create, test, and validate IPs and designs effortlessly, and automatically, whenever possible.
It includes our FPGA shell, several scripts and drivers to facilitate seamless, efficient, and flexible exploitation of the FPGA capabilities by all users, regardless of the specific characteristics of the designs.

\begin{figure}[!t]
\centerline{\includegraphics[width=0.9\linewidth]{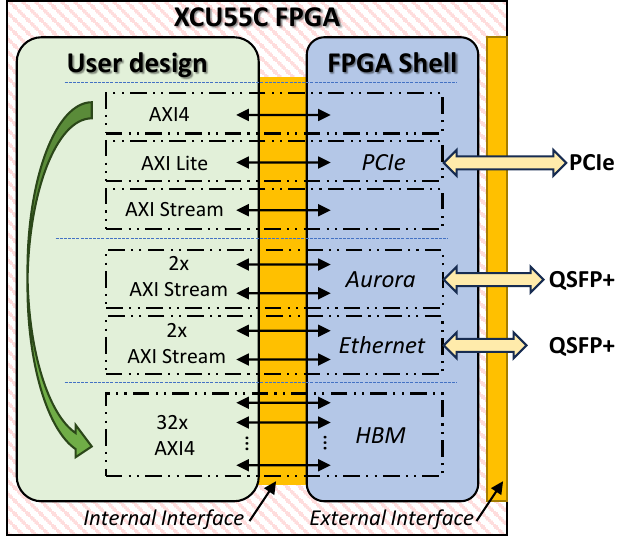}}
\vspace{-0.4cm}
\caption{Diagram of the FGA Shell interfaces}
\label{fig:fig_Shell_DB}
\vspace{-0.4cm}
\end{figure}

The FPGA Shell provides connectivity, through standard protocols, to memory and numerous interfaces allowing seamless communication to the host and other FPGAs. 
Ultimately, our FPGA Shell wrappers 
the user design by parsing the user-specified interfaces to the corresponding FPGA Shell interfaces.
This guarantees that inner designs can be interchangeable while adhering to a defined I/O interface between the FPGA Shell and the design.
Designs range from RISC-V-based implementations to IPs such as neural network implementations.
The broad goal is that users focus only on their designs while abstracting away from issues of designing base components; instead, only well-known interfaces (mainly AXI) need to be provided Fig\ref{fig:fig_Shell_DB}.
Meanwhile, our FPGA Shell takes responsibility for managing the interfaces and making user designs FPGA-agnostic.

Moreover, our FPGA shell is open-source licensed for users to take advantage of its features (if the targeted FPGAs are supported) or want to contribute to its development. 
This represents a paradigm shift in an industry predominantly monopolized by proprietary solutions.
During project creation, our configuration file allows to "enable" and "configure" a combination of IPs including:
\begin{itemize}[align=right,itemindent=1em,labelsep=2pt,labelwidth=1em,leftmargin=*,nosep]
\item Communication with the host CPU and/or other FPGAs via PCIe.
\item Memory access through HBM.
\item Communication with the host CPU and/or other FPGAs via 100 Gb Ethernet.
\item Communication with other FPGAs using Quad Small Form-factor Pluggable (QFSP) and Aurora (Peer-to-Peer communication).
\item Additional modules such as DDR4, UART, a custom BitInfo ROM, or JTAG, can also be provided per user requirements.
\end{itemize}

\begin{table}[!b]
\centering
\caption{FPGA Shell IPs and Features}
\vspace{-0.4cm}
\label{tab:tab_ips_features}
\resizebox{\columnwidth}{!}{%
\begin{tabular}{ll}
\hline
\multicolumn{1}{c}{IPs} & \multicolumn{1}{c}{Features}        \\ \hline
DRAM                    & DDR4,   HBM2 (DMA)                  \\
PCIe                    & Gen4   (qdma)                       \\
Ethernet                & 10Gb/100Gb (over QSFP port)       \\
FPGA-2-FPGA             & MAC-Layer   Aurora (over QSFP port) \\
UART                    & Debug,   Bitstream Uploading        \\
Others                  & JTAG,   InfoROM                     \\ \hline
\end{tabular}%
}
\end{table}

Components shown in Table \ref{tab:tab_ips_features} are designed to operate at maximum capacity while incurring minimal resource costs insignificant in FPGAs with over a million LUTs. 
However, evaluating our FPGA Shell solely in terms of performance would be an incomplete assessment.
Beyond this, our FPGA Shell brings substantial advantages in terms of usability, scalability, and time saved during project setup and debugging.

\subsection{PCIe IP}
Our FPGA Shell uses Peripheral Component Interconnect Express (PCIe) interface to establish communication between the FPGAs and the host server.
PCIe offers advantages in terms of available OS drivers and integrated hard IPs within the FPGA.
The Alveo U55C has PCIe Gen4 with 16x full duplex Lanes running at 8.0 GT/s for a peak bandwidth of 8 GT/s x2 (duplex) x16 lanes / 8 bits/byte = 32 GB/s.
Setting up the PCIe link between the host and FPGA involves two key aspects: (1) Setting up the host by compiling and installing drivers, and (2) setting up the PCIE4C \cite{amd_xilinx_ultrascale_2023-1} hard block inside the FPGA using Vivado, including parameters like BAR options, number of lanes, physical functions, virtual functions, etc.

Xilinx supports two PCIe blocks: XDMA (External Direct Memory Access) and QDMA (Queue Direct Memory Access). 
Our FPGA Shell opts for the more flexible and powerful QDMA implementation, distinguishing itself from standard XDMA found in platforms like Xilinx Vivado.
QDMA key differentiator is the concept of queues, individually configurable by interface type, offering a low-overhead option for continuous update functionality.
Although our FPGA-Shell uses QDMA, our cluster supports PCIe functionalities for bitstreams programmed with QDMA or XDMA host drivers.

Our FPGA-Shell abstracts users from the PCIe IP configuration and setup complexity with this line in the configuration file: 

\begin{lstlisting}[language=html, title=Code1: PCIe configuration]
PCIE,yes,pci_axi,,PCIE_CLK,pcie_clk,pcie_rstn,dma,0,,02
\end{lstlisting}

The first value of this line, which is one of the many in \ding{182} in  Fig.\ref{fig:fig_rapid_prototyping} configuration file, is always the name for the interface to be configured. The second is if the user wants to enable it [yes/no]. The third one is editable and is the name of the AXI interface to be connected to the IP followed by the protocol standard [AXI4-256/AXI4-256/AXI4LITE-64] or blank (like in this PCIe case)	for fixed interfaces. The clocks and reset are also editable and configured next and the last field is for the utilized HBM channel.

\subsection{DRAM: HBM \& DDR4}
High Bandwidth Memory (HBM) is a pivotal component of our FPGA Shell because as the only memory available in the Alveo U55C plays the role of main memory and enables self-hosting capabilities for accelerators. 
HBM uses a 3D-stacked Dynamic Random Access Memory (DRAM) approach to significantly enhance system memory bandwidth through a system-in-package technology.

\begin{figure}[!t]
\centerline{\includegraphics[width=\linewidth]{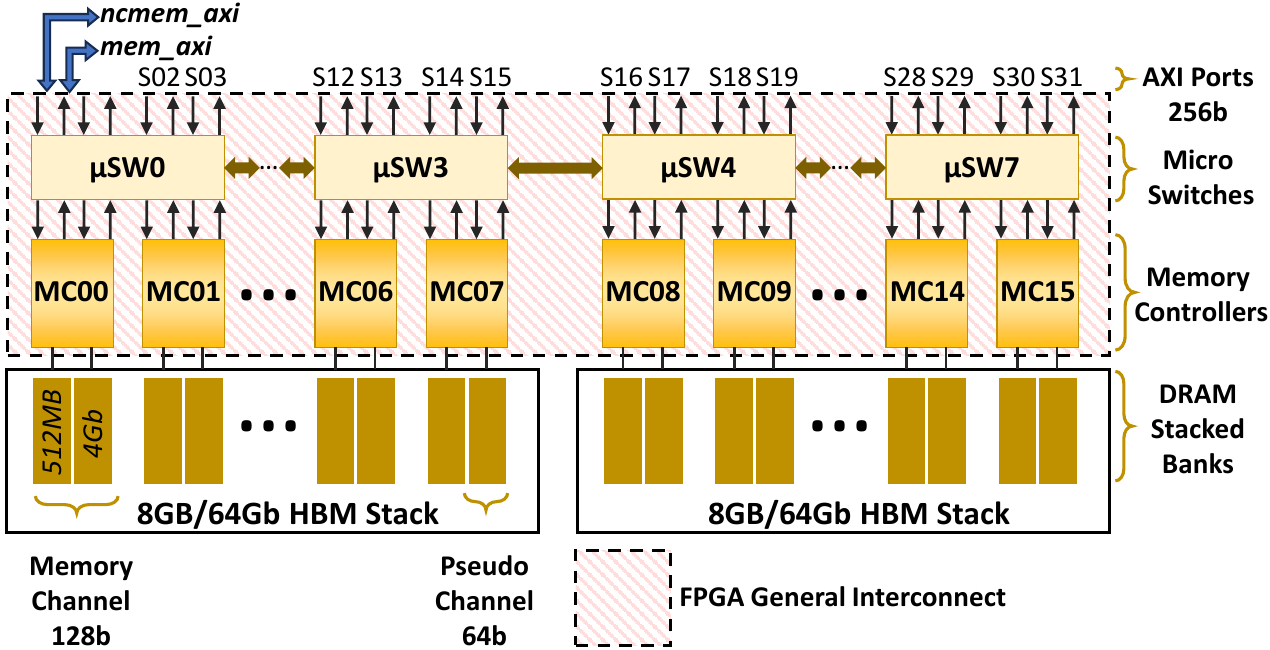}}
\vspace{-0.4cm}
\caption{Interfacing the Alveo U55C HBM (modified from \cite{xilinx_inc_axi_2021})}
\label{fig:fig_hbm_ip_u55}
\end{figure}

Our FPGA shell provides standard interfaces to 
the Alveo U555C HBM controller with two stacks of 64Gb density (8H stack / 8GB) (Fig.\ref{fig:fig_hbm_ip_u55}). 
Each stack has 8 Memory Controllers (MC) and 8 channels divided into 16 pseudo-channels of 256MB.
With a 256-bit AXI Port operating up to 450MHz for each 32 pseudo-channel for a 460.8GB/s bandwidth.
Every 4 AXI Port enters a fully implemented microswitch providing the same behavior for any access inside the 4 pseudo-channels comprised.
The two stacks involve 8 of these micro-switches connected adjacently allowing access to the whole memory.
As HBM consists of multiple SDRAM cores, each controlled by its own MC, our FPGA Shell supports an arbitrary number of AXI-MM channels for connecting designs to multiple MCs.
This feature enables high bandwidth due to the potential increase in concurrency of data exchange in multi-core systems through the distribution of concurrent memory accesses over multiple MCs.

Our FPGA Shell abstracts users from the HBM configuration and setup complexity with these lines in the configuration file: 

\begin{lstlisting}[language=html, title=Code2: HBM configuration]
DDR,no,mem_axi
HBM,yes,mem_axi,AXI3-256,CLK0,0x0,mem_calib_complete,00
HBM,yes,ncmem_axi,AXI3-256,CLK0,0x0,mem_calib_complete,01
HBM,no,mcx_axi0,AXI3-256,CLK0,0x0,mem_calib_complete,04
\end{lstlisting}

In Fig.\ref{fig:fig_hbm_ip_u55} we illustrated how pseudo channels \textit{S00} and \textit{S01} are connected using the AXI3-256 protocol to the defined signals \textit{ncmem\_axi} and \textit{mem\_axi} as explained before. \textit{mcx\_axi0} has everything well configured but is set to \textbf{no} enable it. 

\subsection{Ethernet IP}
Our FPGA Shell Ethernet IP enhances user design capabilities with IP-based connections for user designs.
This extends the concept of self-hosted designs, controlled by built-in CPU(s), which can be RISC-V-based with an OS running on top i.e.
Our Ethernet IP solution satisfies the requirements for standard external connectivity based on CPU needs, mainly offering {Ethernet over QSFP}. For this, we used a set (4-lanes) of FPGA-level Ultrascale+ GTY transceivers providing a peak bandwidth of 32.75 Gb/s per external differential pins pair

This approach facilitates communications between the FPGA-embedded OS and any other Ethernet system via QSFP.
The main originator of Tx data and acceptor of Rx data present in the design is a DMA engine accompanied by its corresponding Tx and Rx memories.
Xilinx AXI DMA core \cite{amd_xilinx_axi_2022} converts AXI4-Stream input/output to AXI4-MM write/read accesses.
\textit{Loopback  FIFOs} implements the feature of data loopback with bypass of the Ethernet core and serves the purpose of checking the test infrastructure itself and debugging the test software.
Finally the design contains all kinds of connections for AXI4-Stream links allowing to debug and characterize data streaming separately for different units.

Our Ethernet-over-PCI driver solution shown in Fig.\ref{fig:fig_ethernet_vec}, provides full connectivity to all 8 FPGA boards within the same node via PCIe, while Ethernet via QSFP establishes connectivity between FPGAs in different (or the same) nodes \cite{castells-rufas_ethernet_2023}.
Additionally, the FPGA Shell provides a homogeneous and scalable communication infrastructure for all FPGAs, irrespective of their design architecture.

The Ethernet IP configuration and setup complexity, at the end, gets hidden from the user behind these lines in our configuration file: 

\begin{lstlisting}[language=Octave, title=Code3: Ethernet configuration]
ETHERNET,yes,eth_axi,AXI4LITE-64,CLK0,100Gb,eth_irq,qsfp1
,hbm,29
\end{lstlisting}

\begin{figure}[!t]
\centerline{\includegraphics[width=0.8\linewidth]{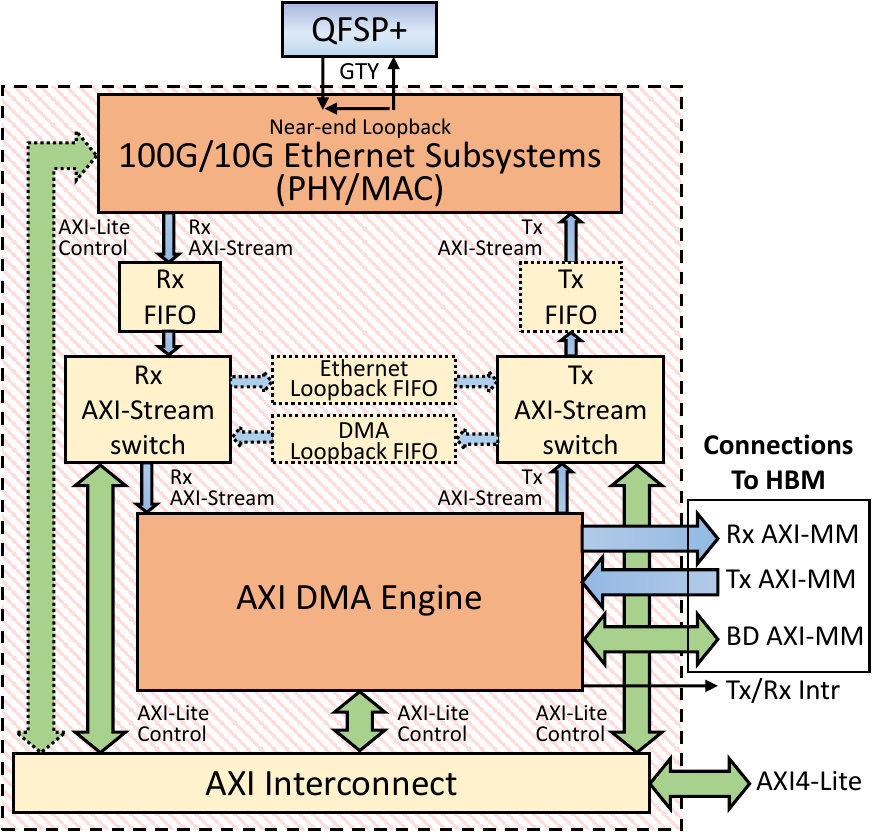}}
\vspace{-0.4cm}
\caption{FPGA Shell: Ethernet IP}
\label{fig:fig_ethernet_vec}
\end{figure}

\subsection{Aurora IP}
Our FPGA Shell seamlessly integrates Xilinx's proprietary lightweight high-speed communication protocol, Aurora.
Aurora is designed to accommodate higher-level protocols, such as TCP/IP, using one or more high-speed serial GT lanes.
It can be referenced to a second or Data Link Layer of the OSI model (the layer where data packets are encoded and decoded into bits).
The FPGA Shell's Aurora subsystem, based on the Aurora 64B/66B IP core, includes SerDes hard macros around FPGA GT differential pin pairs.
The subsystem features receive and transmit 256-bit wide AXI4-Stream channels, along with logic for control, buffering, and diagnostics.
The block diagram of our Aurora solution is quite similar to the Ethernet one in Fig.\ref{fig:fig_ethernet_vec} just replacing the Ethernet subsystems by the Aurora 64B/66B IP because we use the AXI DMA for the same purposes, for providing communications over the IP by software.
The \textit{dotted} blocks are removed from our aurora solution because they are not needed due to a different clock synchronization strategy.

Configured by default to use all 4 GTY lanes connected to an optical QSFP+ connector of Alveo FPGA boards (10 Gb/s per lane), the Aurora subsystem achieves a bandwidth of around 3.5 Gb/s for data exchange between two boards using the Aurora DMA solution under Buildroot Linux with a non-cached region of HBM. 
The developed test application for the FPGA Shell's Aurora subsystem ensures data integrity checks, including ICMP packet exchanges, providing an example of running IP protocol packets over the Aurora link layer.

The Aurora IP configuration and setup complexity, gets hidden from the user behind these lines in our configuration file: 

\begin{lstlisting}[language=Octave, title=Code4: Aurora configuration]
AURORA,no,eth_axi,AXI4LITE-64,CLK0,dma,eth_irq,qsfp1,hbm
,13
\end{lstlisting}

\section{Experimental Results}
\label{sec:analysis}

\begin{figure}[t]
\centerline{\includegraphics[width=\linewidth]{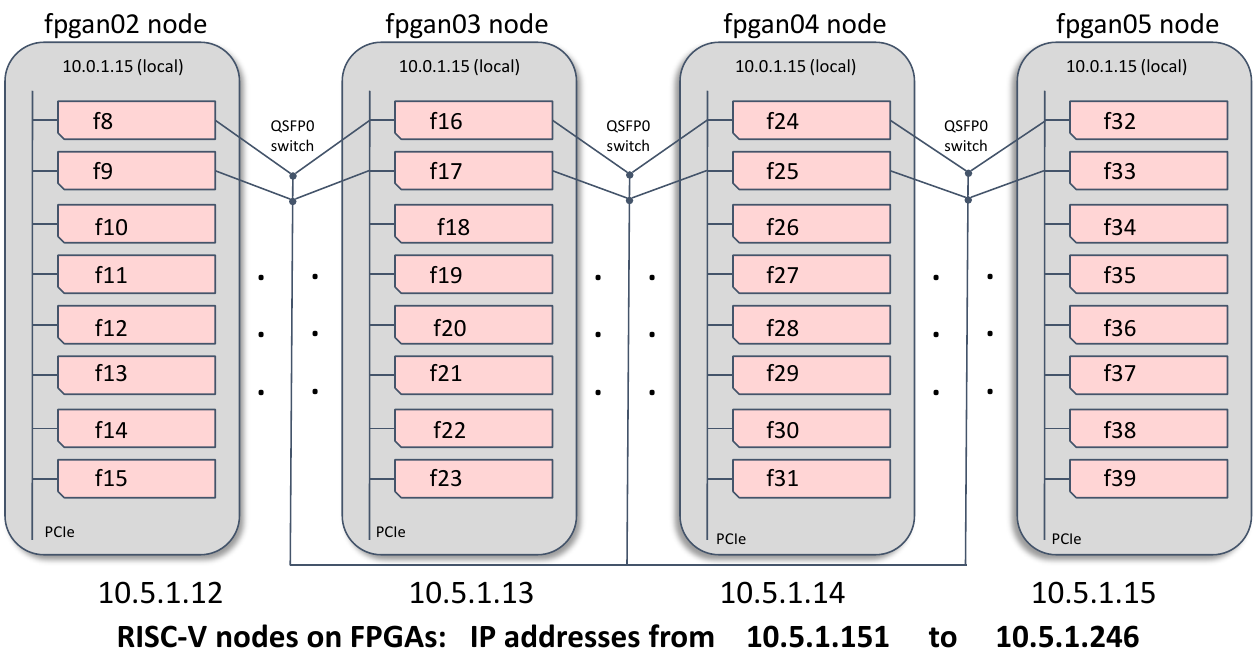}}
\vspace{-0.4cm}
\caption{MPI execution environment}
\label{fig:res_experiments}
\end{figure}

A comprehensive set of bring-up tests was conducted to validate the proper operation of \textit{Our HW Platform} and \textit{Our SW Platform}.
In this section, we delve into a more extensive experiment involving MPI communication across many cores on many FPGAs, leveraging the capabilities of the FPGA Shell and the readiness of our multi-FPGA cluster.
The baseline environment for these experiments is illustrated in Fig.\ref{fig:res_experiments}, utilizing four computing nodes in our cluster; from nodes 2 to 5.

For these experiments, all FPGAs were programmed with an identical bitstream, including a RISC-V microarchitecture with 4 Lagarto Hun CPUs \cite{marenostrum_experimental_exascale_platform_meep__bsc_scalar_nodate}, and the FPGA Shell facilitating interfaces to HBM, PCIe, and Ethernet.
The RISC-V infrastructure on each FPGA is booted to execute Fedora Linux OS on the RISC-V cores. This provides 32 FPGAs able to run MPI applications using the QSFP0 connection to the switch. 

The Fedora distribution image was updated to support booting on our cluster infrastructure in such a way that:
\begin{itemize}[align=right,itemindent=1em,labelsep=2pt,labelwidth=1em,leftmargin=*,nosep]
\item (\textit{Ethernet-over-QSFP}): This Ethernet network is 
dynamically configured by starting the Network Manager service and the \textit{dhclient} servers to get the DHCP configuration by contacting the DHCP server running on our cluster infrastructure.
For this QSFP configuration through DHCP, we propagated the node MAC address from the boot sequence, including both the node and the FPGA numbers, within the binary boot file.
In addition, OpenSBI inserts the proper MAC address into the device tree, for Linux to configure the eth0 device.
We also support through this network the MPI communications being used in these experiments.
\end{itemize}

\begin{figure}[!b]
\vspace{-0.4cm}
\centerline{\includegraphics[width=\linewidth]{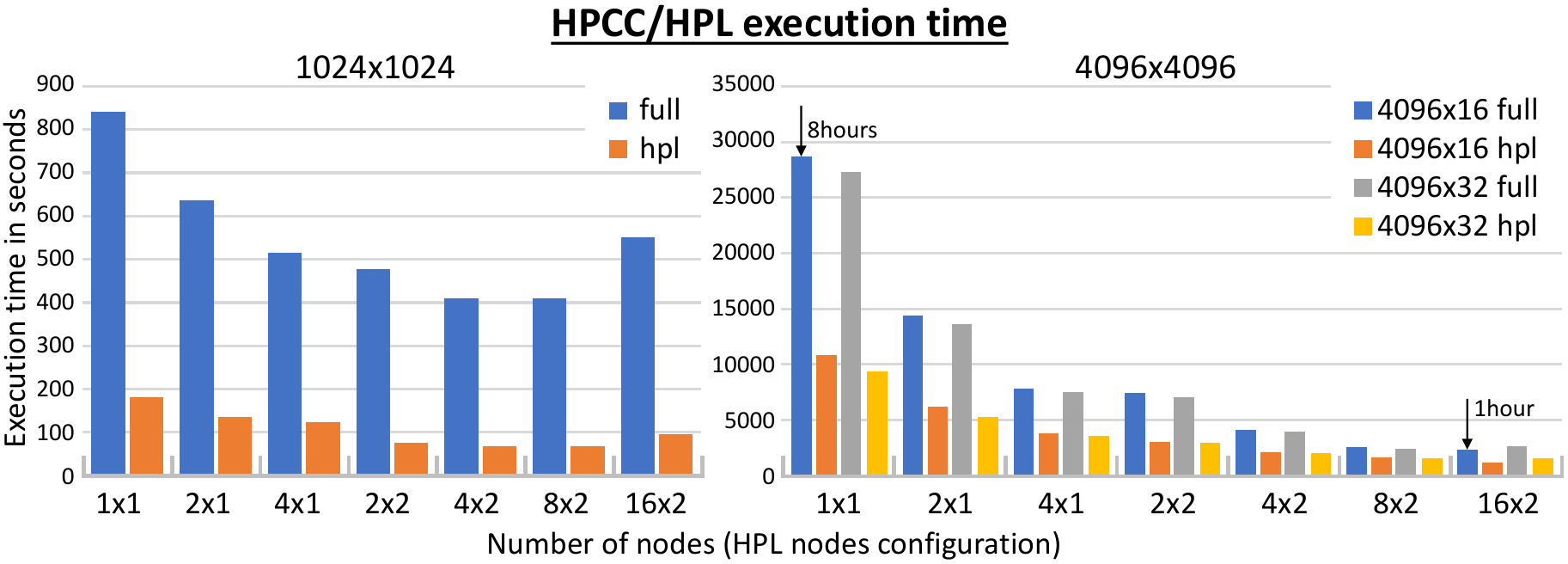}}
\caption{MPI execution results}
\label{fig:res_graphs_vec}
\end{figure}

Fig.~\ref{fig:res_graphs_vec} shows the results obtained for the HPC Challenge application with a different set of FPGAs, increasing the number of FPGAs from 1 to 32, with the configuration of HPCC to be from 1x1 FPGA mesh to 16x2 FPGA mesh.
Both graphics illustrate the Execution Time, showing satisfactory scalability for matrices of 1024x1024 and 4096x4096, even though the times are relatively high, as expected, due to the emulation of RISC-V cores over FPGAs.
The left-side graphic reveals some degree of scalability for both the full application execution time and the particular High-Performance Linpack (HPL) execution time, even with a smaller matrix.
On the right-side graphic, we utilized a larger matrix and tested different block sizes of 16x16 and 32x32, being the latter one faster for every configuration.
In this case, the scalability exhibited is acceptable, decreasing execution time from 8 hours for a single FPGA to 1 hour for 32 FPGAs (configuration 16x2 of HPCC).

\section{Conclusions}

In this paper, we have shown our proposal for the emulation of chip functionality before reaching silicon production, for RISC-V-based designs using our large-scale multi-FPGA cluster.
This platform supports the AMD/Xilinx Alveo U55c FPGA boards, installed on a system with 12 nodes and 8 FPGAs per node. Within the platform, we provide support for various common components, like HBM and DDR4 global memory, PCIe
Gen4, and QSFP Ethernet connections (QSFP0 and QSFP1).
We develop a specific set of common resources, which simplify and automate FPGA-based design generation. 
Moreover, they allow us to run full-fledged Linux distributions on the chips with the possibility to test parallel applications.
As a proof of concept, we have run the HPC Challenge benchmark on a simple experiment with up to 32 FPGAs, showing that the performance of the execution of  the application has an 8-fold improvement, as compared to the execution on a
single node.
These experiments not only validate the correct functioning of all FPGA Shell IPs and the cluster infrastructure but also showcase the potential of deploying multiple designs that can successfully communicate.
This intercommunication can be used to exploit more logic resources and accommodate larger designs for enhancing overall performance.

\begin{acks}
This work has received funding from the MEEP project (European High-Performance Computing Joint Undertaking (JU) under grant agreement No 946002), from ”Lenovo-BSC Contract-Framework (2022)”, from the Spanish Government (PID2019-107255GB-C21/AEI /10.13039/501100011033, CEX2021-001148-S funded by MCIN/AEI / 10.13039/501100011033), and from Generalitat de Catalunya (contract 2021-SGR-01007).

The authors would like to thank John Davis, Luis Plana, Miquel Moretó, Daniel Jiménez Mazure and the Operations Team at BSC for their helpful support and collaboration.
\end{acks}
\bibliographystyle{ACM-Reference-Format}
\bibliography{rapido.bib}


\begin{thebibliography}{29}


\ifx \showCODEN    \undefined \def \showCODEN     #1{\unskip}     \fi
\ifx \showDOI      \undefined \def \showDOI       #1{#1}\fi
\ifx \showISBNx    \undefined \def \showISBNx     #1{\unskip}     \fi
\ifx \showISBNxiii \undefined \def \showISBNxiii  #1{\unskip}     \fi
\ifx \showISSN     \undefined \def \showISSN      #1{\unskip}     \fi
\ifx \showLCCN     \undefined \def \showLCCN      #1{\unskip}     \fi
\ifx \shownote     \undefined \def \shownote      #1{#1}          \fi
\ifx \showarticletitle \undefined \def \showarticletitle #1{#1}   \fi
\ifx \showURL      \undefined \def \showURL       {\relax}        \fi
\providecommand\bibfield[2]{#2}
\providecommand\bibinfo[2]{#2}
\providecommand\natexlab[1]{#1}
\providecommand\showeprint[2][]{arXiv:#2}

\bibitem[{AMD Xilinx}(2022a)]%
        {amd_xilinx_alveo_2022}
\bibfield{author}{\bibinfo{person}{{AMD Xilinx}}.} \bibinfo{year}{2022}\natexlab{a}.
\newblock \bibinfo{booktitle}{\emph{Alveo {U55C} {Data} {Center} {Accelerator} {Cards} {Data} {Sheet}}}.
\newblock \bibinfo{type}{{T}echnical {R}eport} DS978.
\newblock


\bibitem[{AMD Xilinx}(2022b)]%
        {amd_xilinx_axi_2022}
\bibfield{author}{\bibinfo{person}{{AMD Xilinx}}.} \bibinfo{year}{2022}\natexlab{b}.
\newblock \bibinfo{booktitle}{\emph{{AXI} {DMA} v7.1 {LogiCORE} {IP} {Product} {Guide}}}.
\newblock \bibinfo{type}{{T}echnical {R}eport} PG021.
\newblock


\bibitem[{AMD Xilinx}(2023a)]%
        {amd_xilinx_ultrascale_2023}
\bibfield{author}{\bibinfo{person}{{AMD Xilinx}}.} \bibinfo{year}{2023}\natexlab{a}.
\newblock \bibinfo{booktitle}{\emph{{UltraScale} {Architecture} and {Product} {Data} {Sheet}: {Overview}}}.
\newblock \bibinfo{type}{{T}echnical {R}eport} DS890 (v4.4.1).
\newblock


\bibitem[{AMD Xilinx}(2023b)]%
        {amd_xilinx_ultrascale_2023-1}
\bibfield{author}{\bibinfo{person}{{AMD Xilinx}}.} \bibinfo{year}{2023}\natexlab{b}.
\newblock \bibinfo{booktitle}{\emph{{UltraScale}+ {Devices} {Integrated} {Block} for {PCI} {Express} {Product} {Guide}}}.
\newblock \bibinfo{type}{{T}echnical {R}eport} PG213.
\newblock


\bibitem[Babb and {others}(1997)]%
        {babb_logic_1997}
\bibfield{author}{\bibinfo{person}{J. Babb} {and} \bibinfo{person}{{others}}.} \bibinfo{year}{1997}\natexlab{}.
\newblock \showarticletitle{Logic emulation with virtual wires}.
\newblock \bibinfo{journal}{\emph{IEEE Transactions on Computer-Aided Design of Integrated Circuits and Systems}} \bibinfo{volume}{16}, \bibinfo{number}{6} (\bibinfo{year}{1997}), \bibinfo{pages}{609--626}.
\newblock


\bibitem[Bailey(2010)]%
        {bailey_emulator_2010}
\bibfield{author}{\bibinfo{person}{B. Bailey}.} \bibinfo{year}{2010}\natexlab{}.
\newblock \bibinfo{title}{Emulator, accelerator, prototype - what's the difference?}
\newblock
\newblock
\urldef\tempurl%
\url{https://www.eetimes.com/emulator-accelerator-prototype-whats-the-difference/}
\showURL{%
\tempurl}


\bibitem[Borgstrom(2014)]%
        {borgstrom_transaction-based_2014}
\bibfield{author}{\bibinfo{person}{T. Borgstrom}.} \bibinfo{year}{2014}\natexlab{}.
\newblock \bibinfo{title}{Transaction-{Based} {Emulation} {Helps} {Tame} {SoC} {Verification}}.
\newblock
\newblock
\urldef\tempurl%
\url{https://www.electronicdesign.com/technologies/eda/article/21800198/transactionbased-emulation-helps-tame-soc-verification}
\showURL{%
\tempurl}


\bibitem[{Cadence Design Systems©}(2023)]%
        {cadence_design_systems_palladium_2023}
\bibfield{author}{\bibinfo{person}{{Cadence Design Systems©}}.} \bibinfo{year}{2023}\natexlab{}.
\newblock \bibinfo{title}{Palladium {Emulation}}.
\newblock
\newblock
\urldef\tempurl%
\url{https://www.cadence.com/en_US/home/tools/system-design-and-verification/emulation-and-prototyping/palladium.html}
\showURL{%
\tempurl}


\bibitem[Castells-Rufas et~al\mbox{.}(2023)]%
        {castells-rufas_ethernet_2023}
\bibfield{author}{\bibinfo{person}{David Castells-Rufas}, \bibinfo{person}{Xavier Martorell}, \bibinfo{person}{Aleix Roca}, \bibinfo{person}{Alexander Kropotov}, \bibinfo{person}{Xavier Teruel}, \bibinfo{person}{Teresa Cervero}, {and} \bibinfo{person}{John~D. Davis}.} \bibinfo{year}{2023}\natexlab{}.
\newblock \showarticletitle{Ethernet {Emulation} over {PCIe} for {RISC}-{V} {Software} {Development} {Vehicles}}. In \bibinfo{booktitle}{\emph{2023 38th {Conference} on {Design} of {Circuits} and {Integrated} {Systems} ({DCIS})}}. \bibinfo{pages}{1--6}.
\newblock
\urldef\tempurl%
\url{https://doi.org/10.1109/DCIS58620.2023.10335994}
\showDOI{\tempurl}


\bibitem[Compton and Hauck(2002)]%
        {compton_reconfigurable_2002}
\bibfield{author}{\bibinfo{person}{K. Compton} {and} \bibinfo{person}{S. Hauck}.} \bibinfo{year}{2002}\natexlab{}.
\newblock \showarticletitle{Reconfigurable computing: a survey of systems and software}.
\newblock \bibinfo{journal}{\emph{ACM Computing Surveys (csuR)}} \bibinfo{volume}{34}, \bibinfo{number}{2} (\bibinfo{year}{2002}), \bibinfo{pages}{171--210}.
\newblock
\newblock
\shownote{Publisher: ACM New York, NY, USA}.


\bibitem[Cui and {others}(2023)]%
        {cui_risc-v_2023}
\bibfield{author}{\bibinfo{person}{E. Cui} {and} \bibinfo{person}{{others}}.} \bibinfo{year}{2023}\natexlab{}.
\newblock \showarticletitle{RISC-V instruction set architecture extensions: {A} survey}.
\newblock \bibinfo{journal}{\emph{IEEE Access}}  \bibinfo{volume}{11} (\bibinfo{year}{2023}), \bibinfo{pages}{24696--24711}.
\newblock
\newblock
\shownote{Publisher: IEEE}.


\bibitem[Dave and {others}(2006)]%
        {dave_implementing_2006}
\bibfield{author}{\bibinfo{person}{N. Dave} {and} \bibinfo{person}{{others}}.} \bibinfo{year}{2006}\natexlab{}.
\newblock \showarticletitle{Implementing a functional/timing partitioned microprocessor simulator with an {FPGA}}. In \bibinfo{booktitle}{\emph{2nd {Workshop} on {Architecture} {Research} using {FPGA} {Platforms} ({WARFP} 2006)}}.
\newblock


\bibitem[Elsabbagh and {others}(2023)]%
        {elsabbagh_accelerating_2023}
\bibfield{author}{\bibinfo{person}{F. Elsabbagh} {and} \bibinfo{person}{{others}}.} \bibinfo{year}{2023}\natexlab{}.
\newblock \showarticletitle{Accelerating {RTL} {Simulation} with {Hardware}-{Software} {Co}-{Design}}. In \bibinfo{booktitle}{\emph{Symposium on {Microarchitecture} ({MICRO}’23)}}.
\newblock


\bibitem[{FireSim©. Powered by Jekyll \& Minimal Mistakes}(2023)]%
        {firesim_powered_by_jekyll__minimal_mistakes_firesim_2023}
\bibfield{author}{\bibinfo{person}{{FireSim©. Powered by Jekyll \& Minimal Mistakes}}.} \bibinfo{year}{2023}\natexlab{}.
\newblock \bibinfo{title}{{FireSim}}.
\newblock
\newblock
\urldef\tempurl%
\url{https://fires.im/}
\showURL{%
\tempurl}


\bibitem[Hassoun and {others}(2005)]%
        {hassoun_transaction-based_2005}
\bibfield{author}{\bibinfo{person}{S. Hassoun} {and} \bibinfo{person}{{others}}.} \bibinfo{year}{2005}\natexlab{}.
\newblock \showarticletitle{A {Transaction}-{Based} {Unified} {Architecture} for {Simulation} and {Emulation}}.
\newblock \bibinfo{journal}{\emph{Very Large Scale Integration (VLSI) Systems, IEEE Transactions on}}  \bibinfo{volume}{13} (\bibinfo{date}{March} \bibinfo{year}{2005}), \bibinfo{pages}{278 -- 287}.
\newblock


\bibitem[Hauck and Agarwal(1996)]%
        {hauck_software_1996}
\bibfield{author}{\bibinfo{person}{S. Hauck} {and} \bibinfo{person}{A. Agarwal}.} \bibinfo{year}{1996}\natexlab{}.
\newblock \showarticletitle{Software technologies for reconfigurable systems}.
\newblock \bibinfo{journal}{\emph{Northwestern University, Dept. of ECE Technical Report}} (\bibinfo{year}{1996}).
\newblock


\bibitem[Hauck and DeHon(2010)]%
        {hauck_reconfigurable_2010}
\bibfield{author}{\bibinfo{person}{S. Hauck} {and} \bibinfo{person}{A. DeHon}.} \bibinfo{year}{2010}\natexlab{}.
\newblock \bibinfo{booktitle}{\emph{Reconfigurable computing: the theory and practice of {FPGA}-based computation}}.
\newblock \bibinfo{publisher}{Elsevier}.
\newblock


\bibitem[Hosokawa and {others}(2007)]%
        {hosokawa_efficient_2007}
\bibfield{author}{\bibinfo{person}{K. Hosokawa} {and} \bibinfo{person}{{others}}.} \bibinfo{year}{2007}\natexlab{}.
\newblock \showarticletitle{Efficient memory utilization for high-speed {FPGA}-based hardware emulators with {SDRAMs}}.
\newblock \bibinfo{journal}{\emph{IEICE transactions on fundamentals of electronics, communications and computer sciences}} \bibinfo{volume}{90}, \bibinfo{number}{12} (\bibinfo{year}{2007}), \bibinfo{pages}{2810--2817}.
\newblock


\bibitem[Hung and Sun(2018)]%
        {hung_challenges_2018}
\bibfield{author}{\bibinfo{person}{W. Hung} {and} \bibinfo{person}{R. Sun}.} \bibinfo{year}{2018}\natexlab{}.
\newblock \showarticletitle{Challenges in large {FPGA}-based logic emulation systems}. In \bibinfo{booktitle}{\emph{Proceedings of the 2018 {International} {Symposium} on {Physical} {Design}}}. \bibinfo{pages}{26--33}.
\newblock


\bibitem[{LiteX}(2023)]%
        {litex_github_2023}
\bibfield{author}{\bibinfo{person}{{LiteX}}.} \bibinfo{year}{2023}\natexlab{}.
\newblock \showarticletitle{Github:}.
\newblock \bibinfo{journal}{\emph{https://github.com/enjoy-digital/litex}} (\bibinfo{year}{2023}).
\newblock


\bibitem[{Marenostrum Experimental Exascale Platform (MEEP) \& BSC}({[n.\,d.]})]%
        {marenostrum_experimental_exascale_platform_meep__bsc_scalar_nodate}
\bibfield{author}{\bibinfo{person}{{Marenostrum Experimental Exascale Platform (MEEP) \& BSC}}.} \bibinfo{year}{[n.\,d.]}\natexlab{}.
\newblock \bibinfo{title}{Scalar core: {Lagarto} {Hun} improvements {\textbar} {MEEP}}.
\newblock
\newblock
\urldef\tempurl%
\url{https://meep-project.eu/media/news/scalar-core-lagarto-hun-improvements}
\showURL{%
\tempurl}


\bibitem[Mezger and {others}(2022)]%
        {mezger_survey_2022}
\bibfield{author}{\bibinfo{person}{B. Mezger} {and} \bibinfo{person}{{others}}.} \bibinfo{year}{2022}\natexlab{}.
\newblock \showarticletitle{A survey of the {RISC}-{V} architecture software support}.
\newblock \bibinfo{journal}{\emph{IEEE Access}}  \bibinfo{volume}{10} (\bibinfo{year}{2022}), \bibinfo{pages}{51394--51411}.
\newblock
\newblock
\shownote{Publisher: IEEE}.


\bibitem[{RISC-V International®}({[n.\,d.]})]%
        {risc-v_international_specification_nodate}
\bibfield{author}{\bibinfo{person}{{RISC-V International®}}.} \bibinfo{year}{[n.\,d.]}\natexlab{}.
\newblock \bibinfo{title}{Specification {Status} - {Home} - {RISC}-{V} {International}}.
\newblock
\newblock
\urldef\tempurl%
\url{https://wiki.riscv.org/display/HOME/Specification+Status}
\showURL{%
\tempurl}


\bibitem[Rizzatti(2014)]%
        {rizzatti_whats_2014}
\bibfield{author}{\bibinfo{person}{L. Rizzatti}.} \bibinfo{year}{2014}\natexlab{}.
\newblock \showarticletitle{What's {The} {Difference} {Between} {FPGA} {And} {Custom} {Silicon} {Emulators}?}
\newblock \bibinfo{journal}{\emph{Electronic Design}} (\bibinfo{year}{2014}).
\newblock


\bibitem[{Siemens©}(2023)]%
        {siemens_veloce_2023}
\bibfield{author}{\bibinfo{person}{{Siemens©}}.} \bibinfo{year}{2023}\natexlab{}.
\newblock \bibinfo{title}{Veloce {Hardware}-assisted {Verification} {System}}.
\newblock
\newblock
\urldef\tempurl%
\url{https://eda.sw.siemens.com/en-US/ic/veloce/}
\showURL{%
\tempurl}


\bibitem[{Synopsys®}(2023)]%
        {synopsys_haps-100_2023}
\bibfield{author}{\bibinfo{person}{{Synopsys®}}.} \bibinfo{year}{2023}\natexlab{}.
\newblock \bibinfo{title}{{HAPS}-100 – {Prototyping} {\textbar} {Synopsys} {Verification}}.
\newblock
\newblock
\urldef\tempurl%
\url{https://www.synopsys.com/verification/prototyping/haps-100.html}
\showURL{%
\tempurl}


\bibitem[Tessier(2008)]%
        {tessier_multi-fpga_2008}
\bibfield{author}{\bibinfo{person}{R. Tessier}.} \bibinfo{year}{2008}\natexlab{}.
\newblock \showarticletitle{Multi-{FPGA} systems: {Logic} emulation}.
\newblock In \bibinfo{booktitle}{\emph{Reconfigurable {Computing}}}. \bibinfo{pages}{637--669}.
\newblock


\bibitem[Wawrzynek and {others}(2007)]%
        {wawrzynek_ramp_2007}
\bibfield{author}{\bibinfo{person}{J. Wawrzynek} {and} \bibinfo{person}{{others}}.} \bibinfo{year}{2007}\natexlab{}.
\newblock \showarticletitle{{RAMP}: {Research} accelerator for multiple processors}.
\newblock \bibinfo{journal}{\emph{IEEE micro}} \bibinfo{volume}{27}, \bibinfo{number}{2} (\bibinfo{year}{2007}), \bibinfo{pages}{46--57}.
\newblock
\newblock
\shownote{Publisher: IEEE}.


\bibitem[{Xilinx Inc.}(2021)]%
        {xilinx_inc_axi_2021}
\bibfield{author}{\bibinfo{person}{{Xilinx Inc.}}} \bibinfo{year}{2021}\natexlab{}.
\newblock \bibinfo{booktitle}{\emph{{AXI} {High} {Bandwidth} {Memory} {Controller} v1.0 {LogiCORE} {IP} {Product} {Guide}}}.
\newblock \bibinfo{type}{{T}echnical {R}eport} PG276 (v1.0). \bibinfo{institution}{Xilinx Inc.}
\newblock


\end{thebibliography}

\end{document}